\documentclass[conference]{IEEEtran}
\IEEEoverridecommandlockouts
\usepackage{cite}
\usepackage{amsmath,amssymb,amsfonts}
\usepackage{algorithmic}
\usepackage{graphicx}
\usepackage{latex-macros}
\usepackage{textcomp}
\usepackage[caption=false,font=scriptsize]{subfig}

\usepackage{stfloats}
\usepackage{xcolor}
\def\BibTeX{{\rm B\kern-.05em{\sc i\kern-.025em b}\kern-.08em
    T\kern-.1667em\lower.7ex\hbox{E}\kern-.125emX}}

\newtheorem{definition}{Definition}

\newcommand{\vehNum}{N}
\newcommand{\timeNum}{T}
\newcommand{\timeLen}{\delta}
\definecolor{myBlue}{rgb}{0,0,0}

\begin{document}

\title{Measuring Flexibility through Reduction Potential\\
}

\author{\IEEEauthorblockN{Polina Alexeenko}
\IEEEauthorblockA{
\textit{National Renewable Energy Laboratory}\\
Golden, CO, USA \\
polina.alexeenko@nrel.gov}
\and
\IEEEauthorblockN{Matthew Bruchon}
\IEEEauthorblockA{
\textit{National Renewable Energy Laboratory}\\
Golden, CO, USA \\
matthew.bruchon@nrel.gov}
\and
\IEEEauthorblockN{Jesse Bennett}
\IEEEauthorblockA{
\textit{National Renewable Energy Laboratory}\\
Golden, CO, USA \\
jesse.bennett@nrel.gov}
}

\maketitle

\begin{abstract}
While electric vehicles (EVs) often exhibit substantial flexibility, harnessing this flexibility requires precise characterization of its timing and magnitude.
This paper introduces the \textit{reduction potential matrix}, a novel approach to EV load flexibility modeling which is both straightforward to calculate and intuitive to interpret.
This paper demonstrates the approach by quantifying flexibility for two distinct commercial vehicle groups---freight vehicles and transit buses---using simulated charging data from Virginia.
While both groups are found to have substantial flexibility, its properties vary across the groups. 
Naturally, this variability manifests in differences in each group's role as a grid resource.
The paper concludes with a discussion on how system planners, fleet operators, and other stakeholders can use the matrix to assess and leverage EV flexibility.
\end{abstract}


\section{Introduction}

Applications of electric vehicles (EVs) as distributed energy resources are diverse, including energy cost minimization, renewable integration, and power quality regulation \cite{arias2019distribution, sadeghian2022comprehensive}.
While these applications have been subjects of theoretical interest for decades \cite{kempton1997electric, rahman2000role, brooks2002vehicle}, the recent surge in EV penetration has catalyzed practical deployments of EV load control mechanisms \cite{venegas_2021_active}.
Rudimentary load control through pricing has become widespread, with dozens of utilities implementing specialized EV rates \cite{black2024survey}.
More ambitiously, real-time EV load control has been used to minimize total load \cite{alexeenko2023achieving}, regulate frequency \cite{lucas2024participation}, and absorb excess solar generation \cite{ghotge2022use}. 

The potency of electric vehicles as a grid resource depends on their \textit{flexibility}, i.e., the extent to which their charging load can be reshaped. 
Precise definitions of EV load flexibility vary. 
Popular heuristics include deadline (the amount of time available to complete charging) \cite{salah2016deadline}, slack (the difference between the deadline and minimum charging time) \cite{neupane2017generation, lee2019acn}, and the ratio of these quantities \cite{kara_estimating_2015, kumar2013v2g}. 
The main limitation of these heuristics is that they are defined in terms of individual vehicle charging requirements and fail to capture the extent to which flexibility aligns across vehicles. 
In settings where \textit{aggregate} flexibility is paramount (e.g., rate planning, fleet operation, and grid modeling), these heuristics may be insufficient.

The joint flexibility of a group of vehicles is given by their \textit{aggregate flexibility set}: the collection of charging profiles respecting vehicle mobility and grid constraints. 
The flexibility set characterizes all possible load shapes, capturing features such as the achievable range of load at a given time period and the rates at which load can be changed. 
Characterizing the flexibility set can be difficult.
Even in simple settings (e.g., where each vehicle's mobility constraints are linear and the aggregate flexibility set is described by the Minkowski sum of convex polytopes \cite{barot2017concise, trangbaek2012exact}), computation of the aggregate flexibility set is generally intractable \cite{gritzmann1993minkowski, al2024efficient}.

The explicit characterization of aggregate flexibility sets is inherently application agnostic and relevant to arbitrary optimization objectives.
A distinct, though related, stream of literature focuses on \textit{implicit} flexibility characterization, where flexibility is quantified through its capacity to serve specific applications.
Implicit characterizations are less general than explicit ones, but they present more directly the material implications of flexibility for the grid. 
For example, \cite{pertl_equivalent_2019} present a model of instantaneous energy and power capacity  based on a stochastic battery model introduced in \cite{hao2014aggregate}. Although this model captures potential at given points in time, it does not incorporate the deadlines associated with loads and thus provides limited information about capacity at future time periods. 
Addressing this limitation, \cite{schlund_flexability_2020} quantify flexibility using time dependent tuples measuring energy, power, and laxity of the load. Aggregate flexibility is then characterized through the distribution over these tuples. 
Closest in spirit to this paper, \cite{develder_quantifying_2016} characterize flexibility as the amount by which load can be reduced (relative to uncoordinated) and the duration over which that reduction can be sustained. 
The main limitation of this stream of literature is its inability to consider the impacts of grid or other constraints on flexibility. 
Because these methods do not involve joint optimization of vehicle charging, they cannot incorporate constraints that are coupled across vehicles (e.g., grid infrastructure capacity constraints). 

\subsection{Contributions}

This paper introduces the \textit{reduction potential matrix}, a characterization of EV load flexibility which balances the advantages of the explicit and implicit approaches. 
Like the implicit approaches, the matrix admits a straightforward interpretation: the matrix captures how much load can be reduced (relative to uncoordinated charging) at given points in time and how long this reduction can be sustained. 
Like the explicit approaches, the characterization is highly versatile. Critically, this approach can account for arbitrary infrastructural constraints, though the complexity of the computation of the matrix depends on the complexity of the underlying constraints. 
For simple settings, e.g., uni-directional charging without additional grid constraints, the matrix is trivial to compute. 

The reduction potential matrix is applied to data-derived synthetic charge session schedules to quantify the flexibility of two commercial vehicle groups: transit buses and freight vehicles.
The vehicles differ in their mobility and charging patterns, and this diversity naturally gives rise to differences in the vehicles' flexibility.
The matrix successfully captures variation across groups in timing, magnitude, and duration of flexibility.
This variation and the broader implications of the observed flexibility for applications including rate design, fleet operation, and grid planning are discussed.

The remainder of the paper is organized as follows. Section~\ref{sec:model} introduces the system model, defines reduction potential, and presents the reduction potential matrix. Section~\ref{sec:studies} illustrates the application of the matrix to the characterization of the flexibility of two commercial vehicle groups using real-world mobility data. Section~\ref{sec:conc} concludes.

\section{Modeling and Measuring Flexibility} \label{sec:model}

\subsection{System Model and Reduction Potential}

Consider a group of $\vehNum$ vehicles indexed according to $i \in \mathcal{\vehNum} = \left[1, \dots, N \right]$. 
Vehicles are assumed to operate over a discrete time horizon consisting of $\timeNum$ time periods of length $\timeLen$. 
Each vehicle $i$ is assumed to arrive at time $a_i$ at an EV charger with maximum charging rate $r_i$.
Throughout, each vehicle's mobility requirements are assumed to be \textup{feasible} in the sense that the required amount of energy can be delivered within the dwell period given the rate constraint. 
In the absence of load management, vehicles charge in an \textit{uncoordinated} manner. 
\begin{definition}[Uncoordinated load]
In the uncoordinated setting, vehicles charge at their maximum rate from arrival until the satisfaction of their energy requirement. That is, vehicle $i$ has uncoordinated load profile $u_i(\tau)$ where
    \begin{align}
        u_i(\tau) = \begin{cases}
                r_i &   a_i \leq  \tau \leq a_i + \frac{e_i}{r_i} \\
                0 & \text{ else } .
                \end{cases}
    \end{align}
\noindent The uncoordinated load of the set of vehicles $\mathcal{N}$ is then simply the sum over the individual uncoordinated loads, i.e.,
    \begin{align}
        u(\tau) = \sum_{i=1}^N u_i(\tau) .
    \end{align}
\end{definition}

The uncoordinated load profile is often not the unique way to meet mobility needs. 
When vehicle loads can be shifted in time, the uncoordinated load associated with a given time period can be larger than the \textit{minimum load}. 
The minimum load over a $k$-length time interval $\left[t, t+1, \dots, t+k-1 \right]$ is the smallest aggregate load respecting vehicle mobility and grid infrastructure constraints. 
It is the optimal solution of the following optimization problem
\begin{align}
     \mathbf{x}^{t,k} =    \argmin_{\mathbf{x} \in \Rbb^{N \times T}} \sum_{\tau = t}^{t + k - 1} \sum_{i = 1}^N x_{i} (\tau) \text{ subject to } \mathbf{x} \in \mathcal{X},
\end{align}
where  $x_i(\tau) \in \Rbb$ is the charging rate of vehicle $i$ at time $\tau$ and $\mathcal{X}$ is a set encoding grid and mobility constraints. The problem's objective is the minimization of total load (summed over vehicles) across the time intervals of interest $\left[t, t+1, \dots, t+k -1 \right]$. 
Flexibility is characterized through \textit{reduction potential}: the difference between the uncoordinated EV load and the minimum load in a given time frame.
\begin{definition}[Reduction potential]
    Given a set of $k$ time intervals starting at time $t$ and constraint set $\mathcal{X}$, the reduction potential between time intervals  $\left[t, t+1, \dots, t+k-1 \right]$ is 
    \begin{align}
         m_{k,t} = \frac{1}{k} \sum_{\tau = t}^{t+k-1} \left( u(\tau) - x^{t,k}(\tau) \right),
    \end{align}
    where $x^{t,k}(\tau)$ is the minimum aggregate load at time $\tau$:
    \begin{align}
        x^{t,k}(\tau) = \sum_{i=1}^N x^{t,k}_i(\tau).
    \end{align}
\end{definition}

In the most basic setting (i.e., where grid constraints and bi-directional charging are not considered), the constraint set $\mathcal{X}$ is defined by the set of equations
\begin{subequations}\label{eq:uni}
    \begin{align}
       &0 \leq x_i (t) \leq r_i \qquad \forall i \in \mathcal{N}, t \in \left[1, \dots, T \right]\label{eq:uni_rate} \\
       &\sum_{\tau = a_i}^{d_i} x_{i}(\tau) \geq e_i  \qquad \forall i \in \mathcal{N}. \label{eq:uni_energy}
    \end{align}
\end{subequations}
Constraint~\ref{eq:uni_rate} ensures that the vehicle's charging rate is nonnegative and below its maximum rate. Constraint~\ref{eq:uni_energy} ensures that the energy requirement is satisfied within the dwell period.
In this basic setting, the constraints of the optimization problem are separable across vehicles and the problem admits an analytic solution which is trivial to compute.

While equation set~\ref{eq:uni} reflects the constraints typically used for simple models of EV load scheduling, additional constraints can be added to reflect more sophisticated settings.
For example, if the vehicles are served by a single transformer with capacity $C$, the constraint 
\begin{align}
    \sum_{i=1}^N x_i(t) \leq C \quad \forall i \in \mathcal{N}
\end{align}
can be added to encode grid infrastructural limits. 
Similarly, more sophisticated network architectures can be represented through power flow constraints (e.g., those described in \cite{sun2018ev, zhou2020admm}).
Settings with grid constraints are generally more computationally difficult than settings only considering vehicle mobility. 
The added difficulty arises from the fact that when vehicle requirements are coupled through the grid constraints, the resultant optimization problems do not in general admit analytic solutions. 
Furthermore, in the case of non-convex constraints (e.g., when modeling disjunctive charging constraints \cite{sun2016optimal, khonji2018challenges} or AC power flow \cite{andrianesis2020distribution, luo2022temporal}), the optimization problem may be challenging to solve even approximately. 

\subsection{The Reduction Potential Matrix} \label{sec:flex_mat}

Meaningful characterization of EV load flexibility requires comparison of load reduction potential across intervals of varying start times and durations. 
Load reduction across different time frames is encoded through the \textit{reduction potential matrix}.
\begin{definition}[Reduction potential matrix]
Given a set of constraints $\mathcal{X}$, a time horizon $T$, and a maximum delay of $D$ time intervals, the flexibility matrix $M(\mathcal{X}, D, T) \in  \Rbb^{T \times D}$ is a matrix whose $k, t$ entry corresponds to the load reduction potential between times $t$ and $t+k-1$, i.e., 
    \begin{align}
        M(\mathcal{X}, D, T) =
            \begin{bmatrix}
            m_{1,1} & m_{1,2} & \dots & m_{1, T} \\
            m_{2,1} & m_{2,2} & \dots & m_{2, T} \\
            \vdots & & & \vdots \\
            m_{D,1} & m_{D, 2} & \dots & m_{D, T}
            \end{bmatrix}.
    \end{align}
\end{definition}

Row entries of the reduction potential matrix represent the evolution of flexibility over the course of the day, and column entries represent the degree to which flexibility can be sustained. 
To cultivate intuition for the relationship between the matrix entries and the flexibility of the underlying set of EVs, consider several simple examples. 
First, consider the setting where vehicles are completely inflexible, i.e., where $$d_i - a_i = e_i/r_i \quad \forall i \in \mathcal{N}.$$
Here, the minimum load $x^{t,k}_i(\tau)$ is precisely equal to the uncoordinated load $u_i(\tau)$ for any choice of starting time $t$ and duration $k$.
The load reduction potential is zero and the matrix $M(\mathcal{X}, D, T)$ is zero everywhere. 
By contrast, in the setting where each vehicle has at least one time interval of slack (i.e., where  $d_i > a_i + e_i/r_i + \delta$ for all $i \in \mathcal{N}$), vehicles have some degree of flexibility. 
The matrix has a non-zero entry (among other places) in the first row whenever an arrival occurs, i.e., 
\begin{align}
    m_{1,t} > 0 \quad \forall t \in  \left[a_1, \dots, a_N\right].
\end{align}

\textcolor{myBlue}{The matrix captures the extent to which load can be shifted without making assumptions about \textit{where} load will be shifted (beyond the fact that the resultant load must respect vehicle and grid constraints).
The matrix thus captures load shifting potential \textit{prior} to any particular scheduling decision.
Critically, when flexibility is used in one time period, it may be impossible to reduce load in a subsequent time period. 
The extent to which flexibility can be used continuously (e.g., in two adjacent time periods) is captured by the matrix columns.}

\section{Empirical Studies} \label{sec:studies}

\subsection{Vehicle Data}  \label{sec:studies_data}

\begin{figure}
    \centering
      \begin{minipage}{.45\linewidth}
        \subfloat[Freight]{\includegraphics[width=.97\linewidth, trim = 0cm 0cm 1cm 0cm, clip]{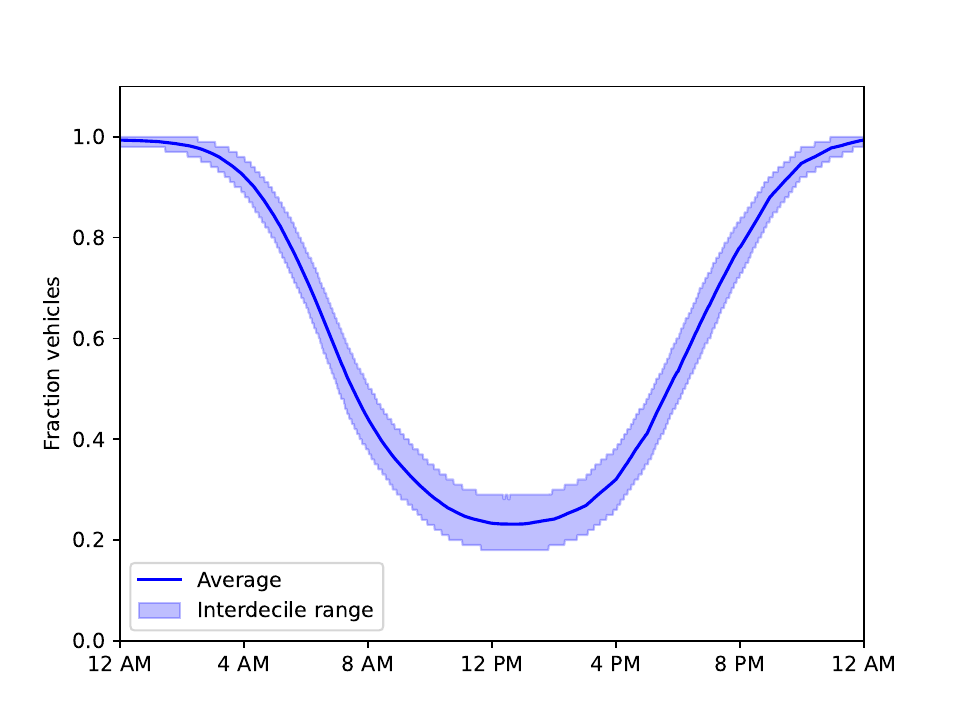}}
    \end{minipage}
    \begin{minipage}{.45\linewidth}
        \subfloat[Transit bus]{\includegraphics[width=.97\linewidth, trim = 0cm 0cm 1cm 0cm, clip]{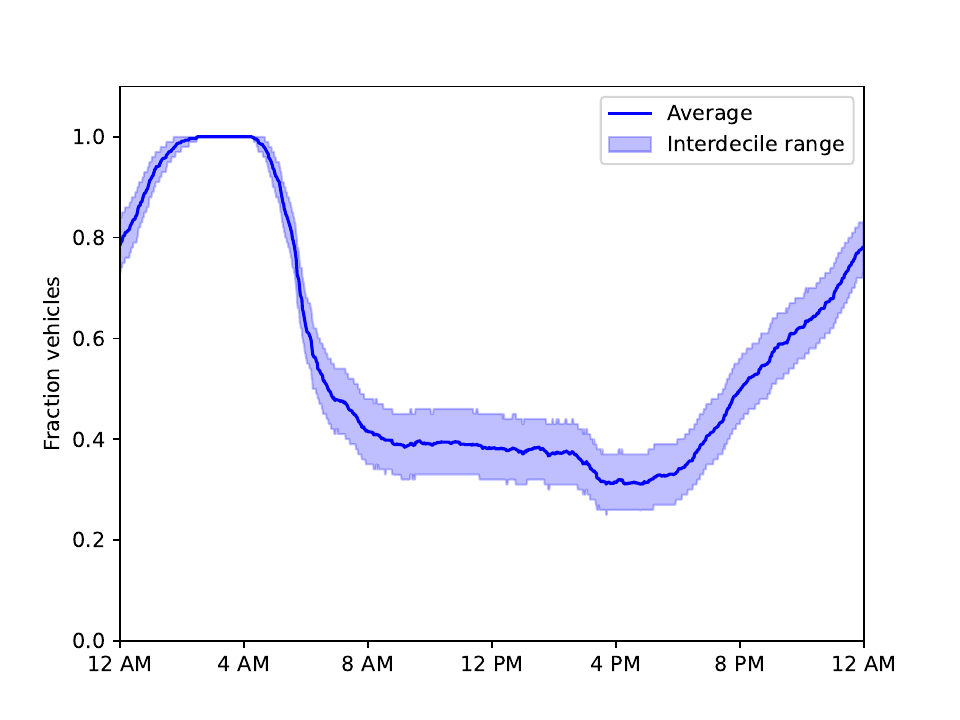}}
    \end{minipage}
    \caption{Empirical dwell probability.}
    \label{fig:dwells}
\end{figure}

\textcolor{myBlue}{The reduction potential matrix is applied to two commercial vehicle vocations (local freight vehicles and transit buses) operating in a high-electrification scenario for Richmond and Newport News, Virginia. 
These vocations are chosen for their dissimilar operations and their potential ease of electrification relative to other vocations of commercial vehicle, such as long-haul trucks. Each vehicles' mobility patterns (travel times, locations, and distances) are synthesized using the Altitude by Geotab telematics data platform as the basis for freight vehicle operations and General Transit Feed Specification data as the basis for transit bus operations. 
These mobility patterns are combined with vehicle and infrastructure specifications (battery sizes and charging rates) anticipated for the year 2040 to model the characteristics of the regions' electrified fleet, assuming around 17,000 local freight vehicles (around 50\% of total) and local freight vehicles and around 500 transit buses (100\% of total) are electrified by 2040 \cite{bruchon2025trb}.}

Charging characteristics and operational patterns differ substantially across the two groups.
Fig.~\ref{fig:dwells} depicts the vehicles' empirical dwell probabilities, i.e., the fraction of vehicles at their domicile throughout the course of the day. 
During periods when the dwell probability is high, it is likely that many vehicles are domiciled simultaneously. 
Notice that while both vehicle groups are connected overnight with high probability, connection timing differs. 
Freight vehicles connect in the early evening while transit buses connect shortly after midnight. 
For both vehicle groups, there is a time of day during which the empirical probability of connection is close to one, i.e., a time during which vehicles are almost certain to be at their domicile. 
For the freight group this period occurs between about 10 PM and 4 AM, while for transit buses the period is shorter, occurring between 1 AM and 5 AM.

\begin{figure}[b]
    \centering
    \begin{minipage}{.45\linewidth}
        \subfloat[Freight]{\includegraphics[width=.97\linewidth]{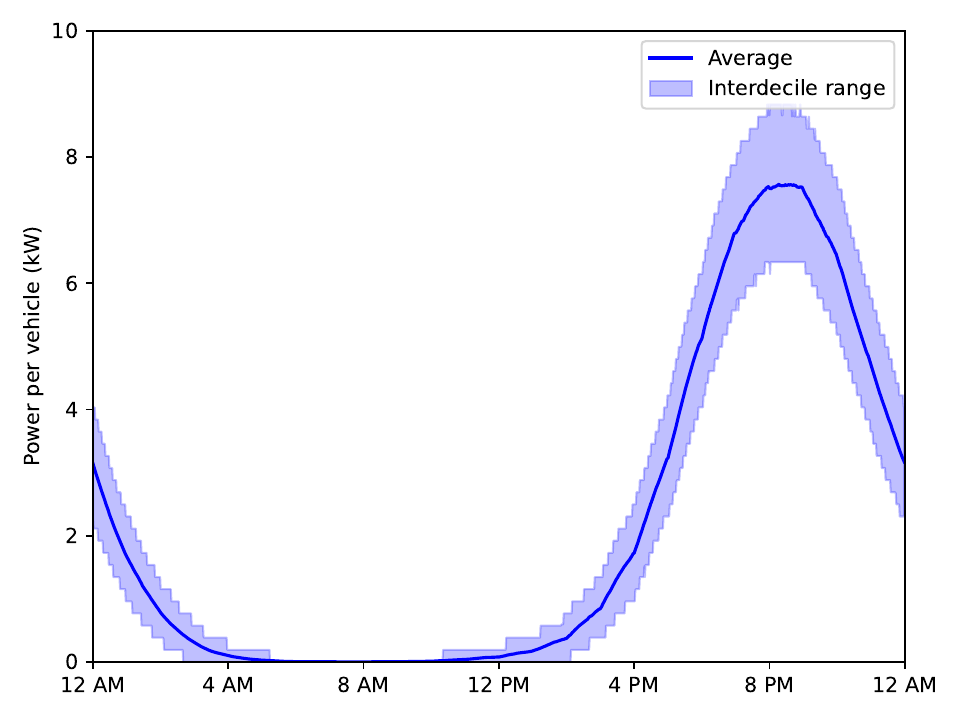}}
    \end{minipage}
    \begin{minipage}{.45\linewidth}
        \subfloat[Transit bus]{\includegraphics[width=.97\linewidth]{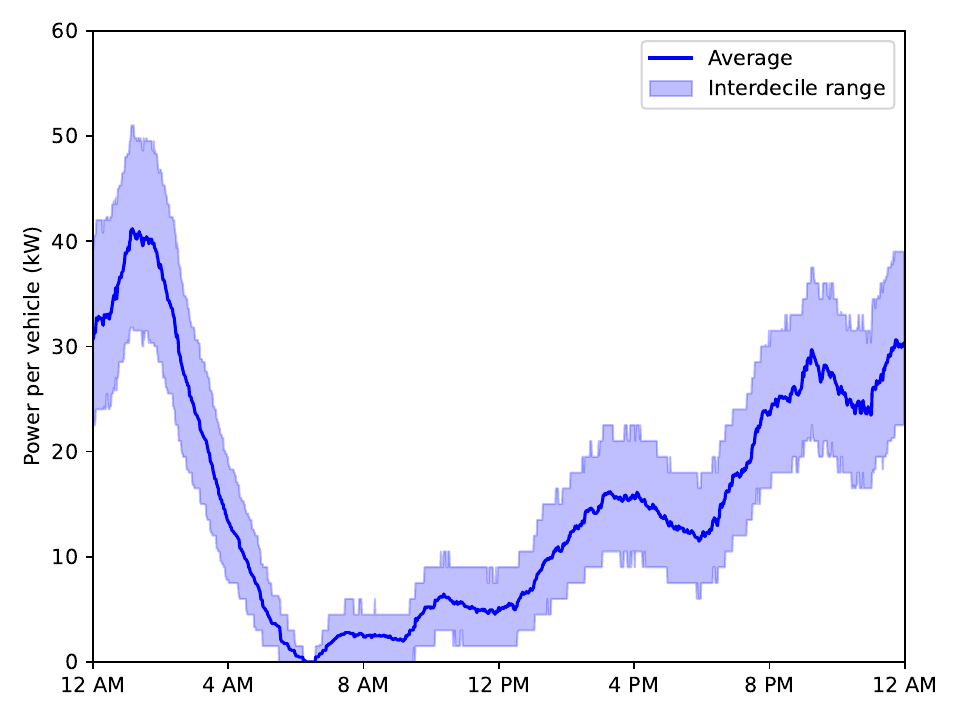}}
    \end{minipage}
    \caption{Average and interquartile range of uncoordinated charging load.}
    \label{fig:unc}
\end{figure}

\begin{figure}[t]
    \centering
      \begin{minipage}{.45\linewidth}
        \subfloat[Freight]{\includegraphics[width=.97\linewidth, trim = 0cm 0cm 0cm 0cm, clip]{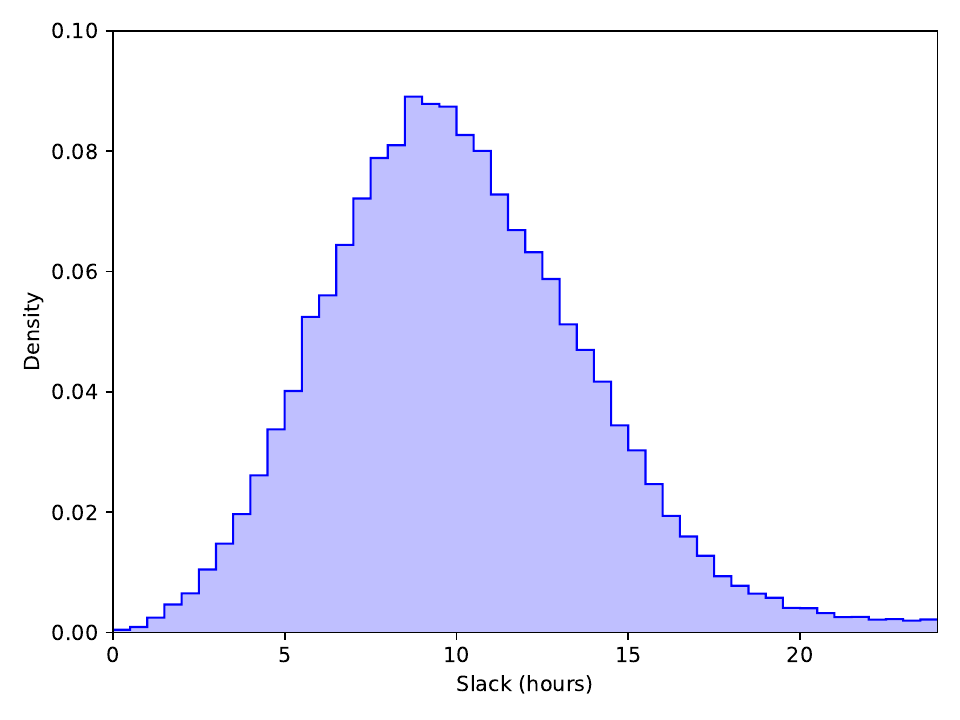}}
    \end{minipage}
    \begin{minipage}{.45\linewidth}
        \subfloat[Transit bus]{\includegraphics[width=.97\linewidth, trim = 0cm 0cm 0cm 0cm, clip]{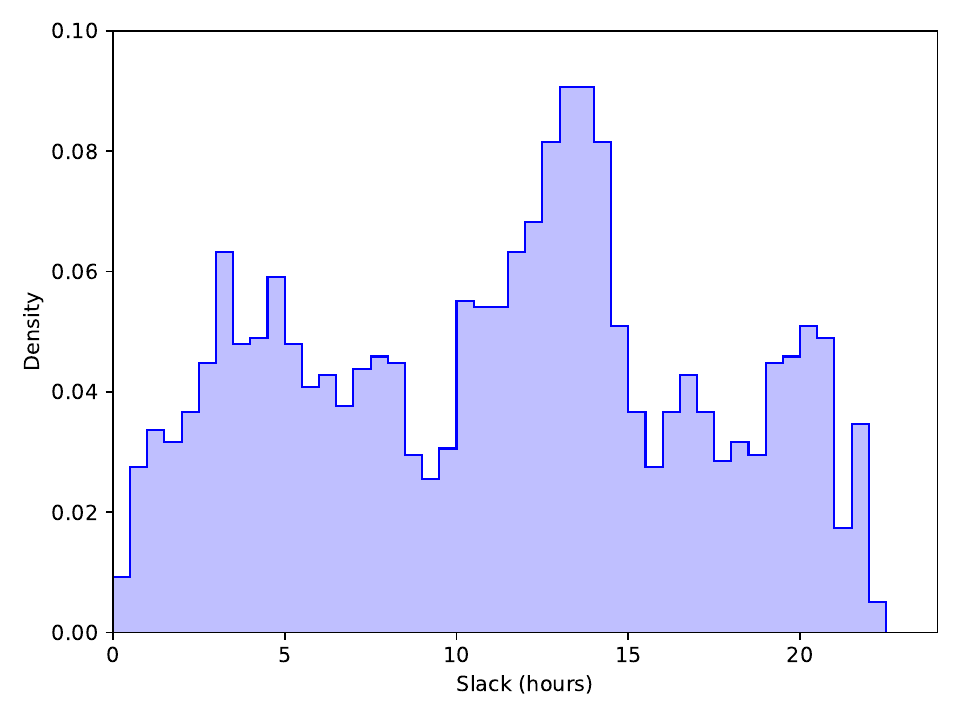}}
    \end{minipage}
    \caption{Slack time distribution.}
    \label{fig:slack}
\end{figure}

Differences in mobility and charging patterns manifest as differences in uncoordinated load across vehicle groups.
Fig.~\ref{fig:unc} depicts the average and interdecile range of daily uncoordinated charging load for each group. 
Peak uncoordinated load tends to coincide with vehicle arrival times, e.g., in the early evening for freight vehicles and around 1 AM for transit buses.
Comparing Figs.~\ref{fig:dwells} and \ref{fig:unc}, note  that vehicle connection times tend to be longer than their uncoordinated charging times. 
\textcolor{myBlue}{This phenomenon is captured in Fig.~\ref{fig:slack}, which shows the empirical distributions over slack times for both vehicle types.
Notice that both groups exhibit substantial slack, having a median time of around 11 hours.}

\subsection{Fleet Flexibility Results} \label{sub_sec:studies}

\begin{figure}[b]
    \centering
    \begin{minipage}{.45\linewidth}
        \subfloat[Freight \label{fig:heat_uni_freight}]{\includegraphics[width=.97\linewidth, trim = 0cm 0cm 1cm 1cm, clip]{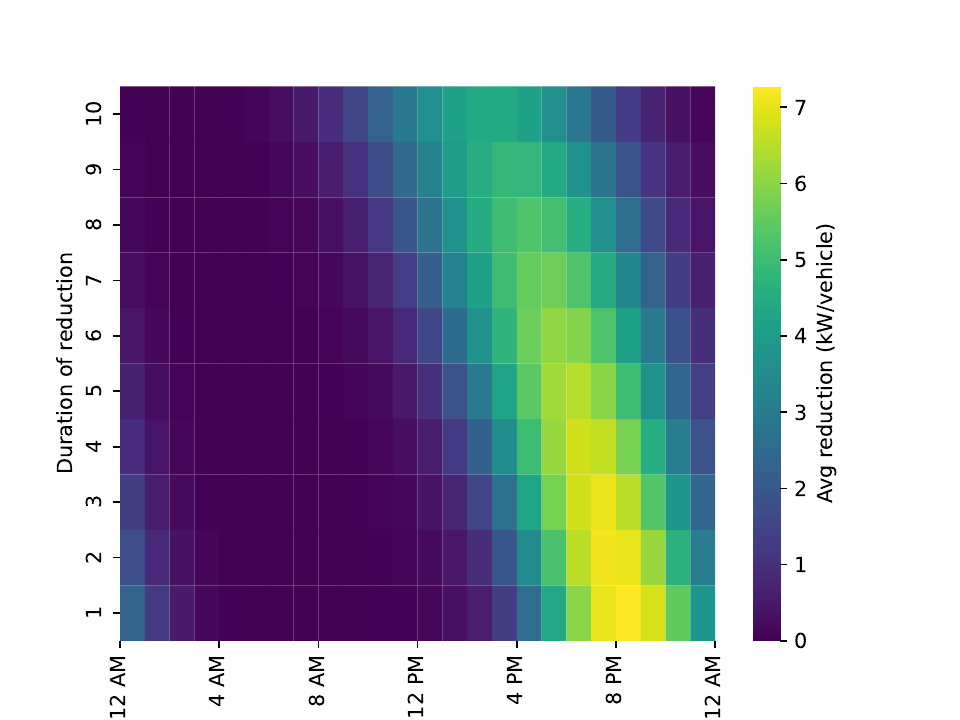}}%
    \end{minipage}
    \begin{minipage}{.45\linewidth}
        \subfloat[Transit]{\includegraphics[width=.97\linewidth, trim = 0cm 0cm 1cm 1cm, clip]{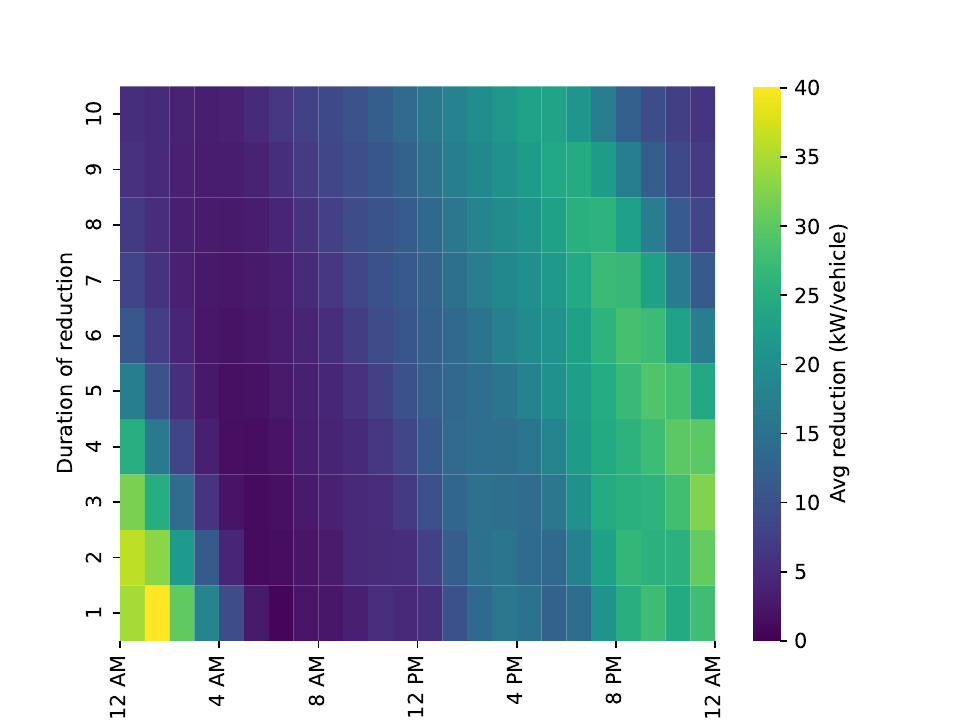}}%
    \end{minipage}
    \caption{Average load reduction potential matrices.}
    \label{fig:heat_uni}
\end{figure}

In this section, the mobility data described in Section~\ref{sec:studies_data} is used to characterize flexibility across the two vehicle groups through the reduction potential matrices developed in Section~\ref{sec:flex_mat}. 
These studies characterize the empirical \textit{average} reduction potential matrices. 
That is, vehicle mobility data is sampled to generate a ``realization" of fleet mobility for a given fleet size. 
The process is repeated to generate a set of samples of reduction potential matrices, and those matrices are added and normalized to obtain an estimate of the average matrix corresponding to the fleet.
Throughout these studies, fleet sizes consisting of one hundred vehicles are considered, and average potential is estimated using one thousand matrix samples. 

Fig.~\ref{fig:heat_uni} shows the average load reduction potential matrix for each group. 
The columns of each heatmap correspond to times of day, while the rows correspond to delay durations. 
Accordingly, the element $t$ columns across and $k$ rows from the bottom captures the amount of load which can be moved outside of the $k$-hour time period starting at time $t$, divided by the number of vehicles $N$ and period duration $k$, averaged over one thousand realizations of hundred-vehicle fleets. 

The bottom row of each heatmap in Fig.~\ref{fig:heat_uni} represents the evolution of one-hour duration load reduction potential throughout the day.
Notice that load reduction potential tends to peak during the times when uncoordinated load peaks for each vehicle group. 
Freight vehicle reduction potential is highest around 8 PM and transit bus reduction potential is highest at 1 AM.
Notice that transit buses have a substantially higher peak reduction potential than do freight vehicles.
This is due largely to the fact that they are associated with higher charging rates and greater energy consumption than freight vehicles.
Note by comparing Figs.~\ref{fig:unc} and \ref{fig:heat_uni} that for both vehicle groups, the peak of the reduction potential is close to the peak uncoordinated load.
This implies that the peak loads can be shifted (by at least one hour) almost in their entirety.

For each vehicle group, there are also times when the reduction potential is close to zero. 
For example, reduction potential is zero-valued whenever there are no vehicles connected.
Furthermore, reduction potential is low-valued when vehicles are connected but unlikely to be charging. 
For example, although freight vehicles disconnect around 7 AM, they tend to finish charging before 4 AM. 
Their one-hour load reduction potential is therefore low-valued between 4 AM and noon, at which point vehicles begin to arrive at the depot.  
\textcolor{myBlue}{\subsection{Applications}}

Load reduction potential matrices can be used for a variety of applications by different stakeholders.
For example, consider a utility wishing to estimate the efficacy of a time-of-use rate with an evening on-peak price and a nighttime off-peak price.
From the results depicted in Fig.~\ref{fig:heat_uni}, the utility can conclude that such a rate may effectively shift freight loads (as these groups have substantial evening-time reduction potential), but would have little impact on transit bus loads, which mainly exhibit reduction potential after midnight.

\textcolor{myBlue}{The matrix can also be used in the identification of candidates for demand response. 
For example, an aggregator wishing to offer load reduction in the evening might seek out a freight fleet rather than a transit fleet. 
More broadly, by comparing the reduction potential matrices of various groups, an aggregator may define different customer segments with different capabilities for load reduction.
Fleet operators themselves may also use the reduction potential matrices. 
In particular, an operator wishing to bid into an energy market as a distributed energy resource can use the matrix to characterize the timing and magnitude of the flexibility they are offering to the grid.
For example, a fleet operator owning 100 transit buses can offer, on average, a load reduction of 4 MW between 1 AM and 2 AM. By contrast, a fleet operator owning 100 freight vehicles can offer, on average, a load reduction of 700 kW between 8 PM and 9 PM. }

\section{Conclusion} \label{sec:conc}

This paper presents the \textit{reduction potential matrix}, a novel approach to capturing the aggregate flexibility of electric vehicle fleets. 
The matrix provides an easily interpretable representation of load flexibility, capturing the timing, magnitude, and duration of load reduction potential. 
The matrix is trivial to compute in settings without grid constraints, but can also be used to quantify flexibility in settings involving infrastructural capacity or network constraints. 
The matrix is applied to charge session data, synthesized from real-world operations data, to characterize the flexibility of two vehicle groups differing in vocation, operation patterns, charging behavior.
This approach successfully captures the variation in the nature of flexibility across the groups, and the implications of this variation for rate design and system planning are discussed.

There are many promising directions for future work. 
For example, it would be interesting to assess whether the matrix can be used to extract information about the operations of individual vehicles in the fleet, and how the temporal granularity of the matrix can be chosen to preserve fleet privacy while providing sufficient information on aggregate flexibility.
Additionally, medium- and heavy-duty vehicle operations and energy needs are substantially more variable than those of light-duty vehicles, but remain relatively underexplored. 
Successful electrification of the medium- and heavy-duty vehicle sector requires broader study of vehicle behaviors across diverse geographies, vocations, and vehicle types.

\section*{Acknowledgment}
This work was authored by the National Renewable Energy Laboratory, operated by the Alliance for Sustainable Energy, LLC, for the U.S. Department of Energy (DOE) under Contract No. DE-AC36-08GO28308. Funding provided by U.S. Department of Energy Office of Energy Efficiency and Renewable Energy Vehicle Technologies Office. The views expressed in the article do not necessarily represent the views of the DOE or the U.S. Government. The U.S. Government retains and the publisher, by accepting the article for publication, acknowledges that the U.S. Government retains a nonexclusive, paid-up, irrevocable, worldwide license to publish or reproduce the published form of this work, or allow others to do so, for U.S. Government purposes.

\bibliographystyle{IEEEtran}
\bibliography{sections/references.bib}

\end{document}